\documentstyle[preprint,prl,aps]{revtex}
\begin {document}

\large
\parindent 0 cm
\begin {center}
{\bf Abelian Cascade Dynamics in Bootstrap Percolation} \\ 
\vskip 0.6 cm
\normalsize
S. S. Manna$^*$
\vskip 0.6 cm
     Satyendra Nath Bose National Centre for Basic Sciences \\
     Block-JD, Sector-III, Bidhannagar,
     Calcutta 700091, India \\
\vskip 0.6 cm
{\underline {Abstract}}
\end {center}
\vskip 0.6 cm
\normalsize

\parindent = 1 cm
\vskip 1.0 cm
The culling process in Bootstrap Percolation is Abelian
since the final stable configuration does not depend
on the details of the updating procedure. An efficient 
algorithm is devised using this idea for the determination 
of the bootstrap percolation threshold in two dimension which 
takes $L^2$ time compared to the $L^3 \log L$ in the 
conventional method.  A generalised Bootstrap Percolation 
allowing many particles at a site is studied where continuous
phase transitions are observed for all values of the threshold
parameter. Similar results are also obtained for the
continuum Bootstrap Percolation model.
\vskip 0.6 cm
\leftline {PACS numbers : 05.50.+q, 64.60.Ak, 64.60.-i, 07.05.Kf}
\vfill
\leftline {$^*$e$-$mail: manna@boson.bose.res.in}
\eject

\parindent=1 cm

In the Bootstrap Percolation model (BPM) sites of a regular 
lattice are occupied randomly with a probability $p$ [1].
A culling condition is imposed: Occupied sites
having fewer than $m$ occupied neighbours are successively 
removed. Finally the lattice may be totally empty or
may have a specific distribution of occupied sites.
Beyond a threshold value $p^*_c(m)$ of $p$
the stable configuration (SC) in an infinitely large system
has a spanning "infinite" cluster with probability 1 [1-3].

Motivation behind studying BPM originates in the properties
of some magnetic materials like $Tb_cY_{1-c}Sb$ where
a strong competition exists between the exchange and crystal
field interactions. Due to the exchange interaction the atomic
spins tend to allign where as the crystal field tend to 
suppress the magnetic moments [4]. In such a pure system where
the effect of exchange interaction is just sufficient to overcome
the effect of the crystal field, the result of introducing
non-magnetic impurities may be significant. A spin with a magnetic
moment if surrounded by too many non-magnetic neighbours
may become non-magnetic. Therefore even at a low temperature, 
as the impurity concentration is gradually increased, at a certain 
stage the magnetic order may get destroyed [1].

In BPM the global connectivity of the stable configuration
is tested with increasing concentration of impurity. By 
definition for all values of $m$, $p^*_c(m) \ge p_c$, the 
ordinary percolation threshold [5].  Applying BP rule on a 
random initial configuration at $p \rightarrow p_c+$, nothing 
is culled for $m=0$, the isolated occupied sites are culled 
for $m=1$ and the dangling chain of sites are also culled 
for $m=2$. In these cases the connectivity remains intact 
which implies $p^*_c(m) = p_c$ for $m$ = 0, 1 and 2 in any 
arbitrary lattice.

A cluster of occupied sites, in which every site has at least
$m$ neighbours, is called an $m$-cluster [2].
The SC contains only $m$-clusters.
An isolated cluster in a $d$ dimensional hypercubic lattice
always has some convex corners on the surface with $d$ neighbours.
Therefore, for $m \ge d+1$ in the hypercubic lattices, 
the SC has only the infinite $m$-clusters.
For example, for $m=3$ BPM on the square lattice,
the SC is a single infinite 3-cluster
with rectangular holes [3]. 
Schonmann has shown that for $m \ge d+1$ in hypercubic lattices
$p^*_c(m)=1$ [6]. This was also predicted from Straley's theory of
`large void instabilities' [7].

The bootstrap percolation probability $Q_m(p)$ is the probability that
a site belongs to the infinite cluster. For those values of $m$ where 
$p^*_c(m)=p_c$, the transition is continuous like percolation.
For other values of $m \ge d+1$ first order transitions are
observed. As $p \rightarrow p^*_c(m)+$, a SC on 
further removal of even a single site finally becomes empty showing a 
discontinuous drop in $Q_m(p)$ indicating a first 
order transition.  

Numerical estimation of $p^*_c(m)$ turns out to be a difficult task.
For a system of size $L$ an effective threshold $p^*_c(m,L)$ is
obtained after configuration averaging which slowly converges to 
$p^*_c(m)$ in the limit of $L \rightarrow \infty$. For example,
for $m=d+1$ on hypercubic lattices, Schonmann has shown that
the asymptotic value is approached with a scaling correction 
$1/{\log}^{d-1} L$ [6]. Using the definition, BPM is best studied
by the "multi-spin coding technique" where occupied and vacant 
sites are represented by the two states of a bit [8-9]. 
A $B$ bit integer word storing $B$ sites is updated by a logical 
equation involving the neighbouring words.  For example, for $m=3$ 
BPM on the square lattice, if the four neighbouring words are $w_1, 
w_2, w_3$ and $w_4$, the central word $w_c$ is updated as:
\[
w_c =not [(\overline w_1+\overline w_2)*(\overline w_3+\overline w_4)+ 
\overline w_1*\overline w_2+
\overline w_3 * \overline w_4]
\]
where $\overline w=not(w)$ and $+$ and $*$ symbols denote the $or$ and 
$and$ operations. The average number of iteration steps necessary to 
reach the SC at $p^*_c$ increases linearly with $L$. Again, the 
$p^*_c$ is estimated by successively halving certain interval $p^{high}$ 
and $p^{low}$ which needs around $ \log L$ number of different runs. 
Therefore
ordinarily the computer effort increases as $L^3 \log L$ near $p^*_c$ for
a two dimensional system of size $L$.

   We first observe that given a particular initial configuration
the SC in any BPM is always the same irrespective
of whether a parallel, sequential or a mixed updating procedure is used.
The `occupation number configuration'(ONC) is described by defining
at every site $i$, $s_i$ = 1 (occupied) or 0 (vacant) and the number of occupied 
neighbours as the `neighbour number'(NN) $n_i = \Sigma^{nn}_j s_j$ 
where $j$ runs over the nearest neighbours.
Then the culling of a site can be written as:
\[
{\rm If}\quad\quad n_i < m, \quad{\rm then}\quad s_i \rightarrow 0 , 
\quad n_i \rightarrow n_o
\]
\[
{\rm and} \quad\quad n_j \rightarrow n_j-1 
\]
where $j$ runs over the unculled occupied neighbours
only and $n_o$ is any fixed large number.
Now consider two arbitrary sites $i$ and $j$ which are 
to be culled. If we first cull 
the site $i$ and then the site $j$, then for some sites
the NNs decrease by one and for the
sites which are common neighbours of both $i$ and $j$ the NNs
decrease by two. We see that the new ONC and NNC 
are exactly the same if the sites $i$ and $j$
were culled in the reverse order. Therefore
the resulting configuration is symmetric under the 
exchange of the sites $i$ and $j$. If we apply this
argument repeatedly we see that at the end of the BP process
the same ONC and NNC appear irrespective
of the choice of sequence in which the different sites are culled.
This process is reminiscent of the Abelian sandpile model (ASM)
of Self-Organized Criticality where same SC
is obtained irrespective of the sequence of the sand grain additions [10].
Hence we shall call our culling process as Abelian.

   A cascade of culling sites can be grown from a single culled site. 
A site $i$ with $n_i < m$ is selected on an arbitrary
ON configuration and culled. It may result in some more sites 
with $n_j < m$ in the neighbourhood and are culled again.
This creates other sites in the further neighbourhood ready for culling.
In this way a cascade of culling sites continues which stops
when no further site can be culled. 
We call the set of sites culled in the cascade
as the `culled cluster' (CC) analogous to the `avalanche cluster' in ASM [11].

   This idea greatly helps us to find out the SC.
The difference between an ON configuration and its SC
is a set of clusters of culled sites. Each such cluster is in general
a combination of many CCs. Since the culling process is Abelian
it does not matter in which sequence these CCs are culled.
Therefore given an ON configuration, we try to grow CCs from all sites 
sequentially one after the other. The remaining unculled 
sites constitute the SC. Since a site can be culled 
at most once, the computer effort is proportional to the area $L^2$ of 
the system compared to $L^3$ in the ordinary method.

   The BP threshold $p^*_c(m)$ can also be obtained starting from a fully
occupied lattice and then deleting the sites one after another.
After deleting a site the corresponding SC is
checked for percolation. The process is continued till the 
SC ceases to percolate. This process however becomes
much easier on using the Abelian property. The 
SC after deleting $(n+1)$ sites is obtained by removing the $(n+1)$-th site from
the SC after deleting $n$ sites and then growing a CC
from the deleted $(n+1)$-th site.
If $f^*_c$ is the fraction of deleted sites when the SC just stops
percolating, then $p^*_c = 1-f^*_c$ is the BP threshold for
this configuration (Fig.1).
Therefore our algorithm takes
$L^2$ time compared to $L^3 \log L$ in the ordinary method.
In ordinary method, huge time is wasted in repeated update attempts
to already culled sites. In our algorithm, only the culling sites
are updated by growing one CC after another. Another method, called
"rectangle" algorithm has some indication to be faster than ordinary
BP algorithm for large systems and near the percolation threshold [12].

   Here we introduce a generalised BPM by allowing a site to be occupied
up by more than one particle. On a square lattice, sites
are filled up randomly to an average density of $\rho$ particles per site
so that $s_i$ is the number of particles at the site $i$.
Here the neighbourhood includes the site itself and NN is
defined as $n_i=\Sigma^{nn}_j s_j+s_i-1$. Now, if $n_i<m$, we delete 
all particles
at $i$ as: $s_i \rightarrow 0$ and the NN at $i$ is raised to
$n_i \rightarrow n_o$ whereas the NNs of the neighbouring unculled sites
are decreased as $n_j \rightarrow n_j-s_i$.
To obtain the SC, CCs
are grown sequentially from all the lattice sites satisfying the
culling condition $n_i < m$. We use the Hoshen   
Kopelman algorithm [13] to test if the SC
has an infinite percolation cluster.

   The percolation thresold $\rho^*_c(m)$ is the minimum 
value of the average initial density $\rho$ beyond which an infinite 
cluster exists with probability 1 in the SC. 
The percolation probability $Q_m(\rho)$ is the fraction of particles
that survive on the infinite percolation cluster. Like BPM on the Cayley 
tree [1], our results indicate a gap as:
$
Q_m(\rho)-Q_m(\rho^*_c(m)) \sim (\rho-\rho^*_c(m))^{\beta}
$
where,
$
Q_m(\rho^*_c(m)) = 
Lim_{\rho \rightarrow \rho^*_c(m)} Q_m(\rho)
$
in the limit $L \rightarrow \infty$.
However, according to the scaling theory [5], for finite systems
the scaling variable should be $L/\xi$, where
$\xi=(\rho-\rho^*_c(m))^{-\nu}$ is the correlation length.
Consequently the percolation probability should behave as:
$
Q_m(\rho)-Q_m(\rho^*_c(m)) = L^{-\beta/\nu}F[(\rho-\rho^*_c(m))L^{1/\nu}]
$
where the scaling function $F(x) \rightarrow x^{\beta}$ for large $L$.
For $m$=2 we calculate the $Q_2(\rho)$ as a function of $\rho$
on system sizes up to $L$ =1024. The curves for different systems are 
found to be continuous and pass through a fixed point with coordinates 
$(\rho^*_c(2),Q_2(\rho^*_c(2))) = (0.899, 0.179)$ (Fig. 2). In Fig. 3, 
$(Q_2(\rho)-Q_2(\rho^*_c(2)))L^{0.075}$ is plotted with 
$(\rho-\rho^*_c(2))L^{0.75}$ and an excellent collapse 
of the data on a continuous curve is observed. We conclude that 
$\nu \approx 4/3$ and $\beta \approx 0.10$ compared to the 
percolation values $4/3$ and $5/36$ in the ordinary percolation [5].

   Since the number of particles at a site is not bounded  
therefore even the isolated sites with $s_i > m+1$ cannot
be culled. Similarly, a pair of adjacent sites $i$ and $j$ 
with $s_i+s_j \ge m+1$ also survives. Thus the SC
at any $\rho$ contains many islands. As we
increase $\rho$, size of these islands increase and they
merge into one another. This merging process is same as in 
ordinary percolation, hence we expect the transition to be
continuous, and in the same universality class.
 
   Suppose now that every sites has exactly $\rho$ particles.
Then $\rho^*_c(m)=(m+1)/5$. However, in a random process a
site has on the average $\rho$ particles with a fluctuation
around $\sqrt \rho$. Therefore even if $\rho = (m+1)/5$ 
half of the sites are culled in the
first iteration. This implies, that to obtain a spanning cluster
of unculled sites in the SC, the 
$\rho^*_c(m)=c_1m+c_2$ where $c_1>1/5$ and $c_2=1/5$.
The variation of $\rho_c(m)$ with $m$ is studied 
and shown in Fig. 4 where $c_1$ = 0.271 and $c_2$
= 0.198 are obtained.

   We finally study a continuum 
BPM in terms of percolation of overlaping boxes [14]. In a region of 
two dimensional continuous space of size $L \times L$ with periodic boundary 
conditions along both the directions we uniformly distribute $N$ points
so that the probability of a point being within the area $dA$ is 
proportional to $dA$. Every point is situated at the centre
of a square box of size $R=1/2$ and if two boxes overlap then the
two points belong to the same cluster. The neighbour number $n_i$ 
of the point $i$ is the number of points within the associated box
including the point itself. All points with $n_i < m$ are culled
and the culling process is repeated till a
SC is reached which cannot be reduced any further (fig. 5).

   To calculate the NNs using a local 
search method we consider an underlying square lattice of size 
$L$. A primitive cell of this lattice is referred by
its bottom-left integer coordinates $(a,b)$. The serial numbers of all points 
$N_{ab}$ within the cell $(a,b)$ are stored in an auxiliary
array $A(a,b,k)$ with $k=1,N_{ab}$ and $A(a,b,0)=N_{ab}$.
For a point $i$ within the cell $(a,b)$, the 
nine cells from $a-1$ to $a+1$ and from $b-1$ to $b+1$
are searched to pick up $n_i$ neighbouring points.
In a second auxiliary array $neb(N,M)$ the serial numbers 
of $n_i$ points are stored in the locations $neb(i,k),k=1,n_i$
and $neb(i,0)$ stores $n_i$.

   Culling clusters are initiated from all sites with $neb(i,0) < m$.
When a point $i$ is culled, we update $neb(i,0) = n_o$, unculled 
neighbours $j$ of $i$ are updated as $neb(j,0)=neb(j,0)-1$ and 
those with $neb(j,0) < m$ are culled in the next time step. 
A SC is shown in Fig. 5. 

   The percolation threshold $\rho^*_c(m)$ and probability $Q_m(\rho)$ are
similarly defined w.r.t. the average initial density $\rho$ of points. Here 
again we find a gap in $Q_m(\rho)$. Variation of $Q_m(\rho)$ is continuous
and for different system sizes, curves pass through a fixed point.
We studied the case of $m$=6, and the fixed point is found to be around (10.55, 4.65).
For this model we could go only up to the system size $L$ = 160 and
$\nu \approx$ 1.4 and $\beta \approx 0.05$ are obtained. 
We believe that this continuum bootstrap percolation percolation model
belongs to the same universality class as the lattice bootstrap
percolation defined above with multiple particles allowed per site, and 
both of these are in same universality class as the ordinary percolation.

   To summarize, we observed that the sequence of culling sites in the
BPM is irrelevant and the same stable configuration is obtained irrespective
of the particular culling sequence used. This knowledge helps us to
devise an efficient algorithm to find out the percolation threshold.
We define a generalization of the bootsrap percolation model which allows
lattice sites to be occupied by more than one particle. 
This is the discretization of the natural continuum version of
percolation. Unlike the earlier studied bootstrap percolation models
which undergo a first order transion for some $m$, we argued, and presented
numerical evidence that the percolation transition for this model is 
continuous for all $m$, and is in the same universality class of usual percolation.

   We acknowledge with thanks D. Dhar for many useful discussions
and suggestions. I also thankfully acknowledge D. Stauffer, J. Adler 
and J. R. Banavar for the critical reading of the manuscript.

\vfill
\eject
\parindent= 0 cm
{\underline {\bf Figure Captions:}}
\vskip 1.0 cm
     
\begin {itemize}

\item [1.] Starting from a fully occupied square 
lattice of size $L$ = 32 with p. b. c., 77 sites (shown by filled 
squares) are deleted one 
after the other. Culled clusters of $m=3$ BPM are grown from 
the deleted sites. The paths of the propagation of the cascading
process is shown by lines. The 77-th cluster had grown from 
the encircled site which evacuates the whole system.
This implies that the $p^*_c(3)$ = (1-77/1024) $\approx$ 0.925
for this configuration.

\item [2.] The percolation probability $Q_2(\rho)$ for the generalised BPM 
with $m$= 2 is plotted with the average density
$\rho$ for system sizes $L$ = 256 (circle), 512 (square) and 1024 (triangles).
The common point of the three curves is 
$(\rho^*_c(2),Q_2(\rho^*_c(2)))$=(0.899, 0.179).

\item [3.] The collapse of $(Q_2(\rho)-Q_2(\rho^*_c(2)))L^{0.075}$
as a function of $(\rho-\rho^*_c(2))L^{0.75}$ is shown using same
symbols for the same system sizes as in Fig. 2. 

\item [4.] For the generalised BPM, the percolation threshold
$\rho_c(m)$ is plotted for various $m$ values. The continuous line
is a linear fit with slope 0.271 and intercept 0.198.

\item [5.] The stable configuration of continuum BPM with $m=3$
using the initial density of $\rho$ = 6.12 points per unit area
in a system of size $20 \times 20$. Every point is at the centre of
a box of size $R=1/2$. The boxes belonging to the infinite cluster
are shown by filled squares where as those of isolated clusters
are shown by open squares.

\end {itemize}
\vfill
\eject
\parindent= 0 cm
{\underline {\bf References:}}
\vskip 1.0 cm
     
\begin {itemize}

\item [1.] J. Chalupa, P. L. Leath and G. R. Reich, J. Phys. C., {\bf 12}, L31 (1979).

\item [2.] P. M. Kogut anf P. L. Leath, {\bf 12}, 3187 (1981).

\item [3.] J. Adler, Physica A, {\bf 171}, 453 (1991),
J. Adler and A. Aharony, J. Phys. A. {\bf 21} 1387 (1988),
J. A. M. S. Duarte, Physica A, {\bf 158}, 1075 (1989).

\item [4.] M. Pollak and I. Riess, Phys. Status Solidi {\bf B69}, K15 (1975).

\item [5.] D. Stauffer, {\it Introduction to Percolation
Theory}, (Taylor \& Francis, London), 1985.

\item [6.] R. H. Schonmann, J. Stat. Phys. {\bf 58}, 1239 (1990).

\item [7.] J. Straley, unpublished results, presented in [2].

\item [8.] J. Adler, D. Stauffer and A. Aharony, J. Phys. A. {\bf 22}
L297 (1989).

\item [9.] S. S. Manna, D. Stauffer and D. W. Heermann, Physica A
{\bf 162} 20 (1989).

\item [10.] D. Dhar, Phys. Rev. Lett. {\bf 64}, 1613 (1990).

\item [11.] S. S. Manna, J. Stat. Phys. {\bf 59}, 509 (1990).

\item [12.] J. Adler, R. Gross and R. Warmund, Physica A {\bf 163}, 440
(1990).

\item [13.] J. Hoshen and R. Kopelman, Phys. Rev. B., {\bf 14}, 3428 (1976)

\item [14.] R. Meester and R. Roy, {\it Continuum Percolation},
Cambridge University Press, 1996.

\end {itemize}
\end {document}
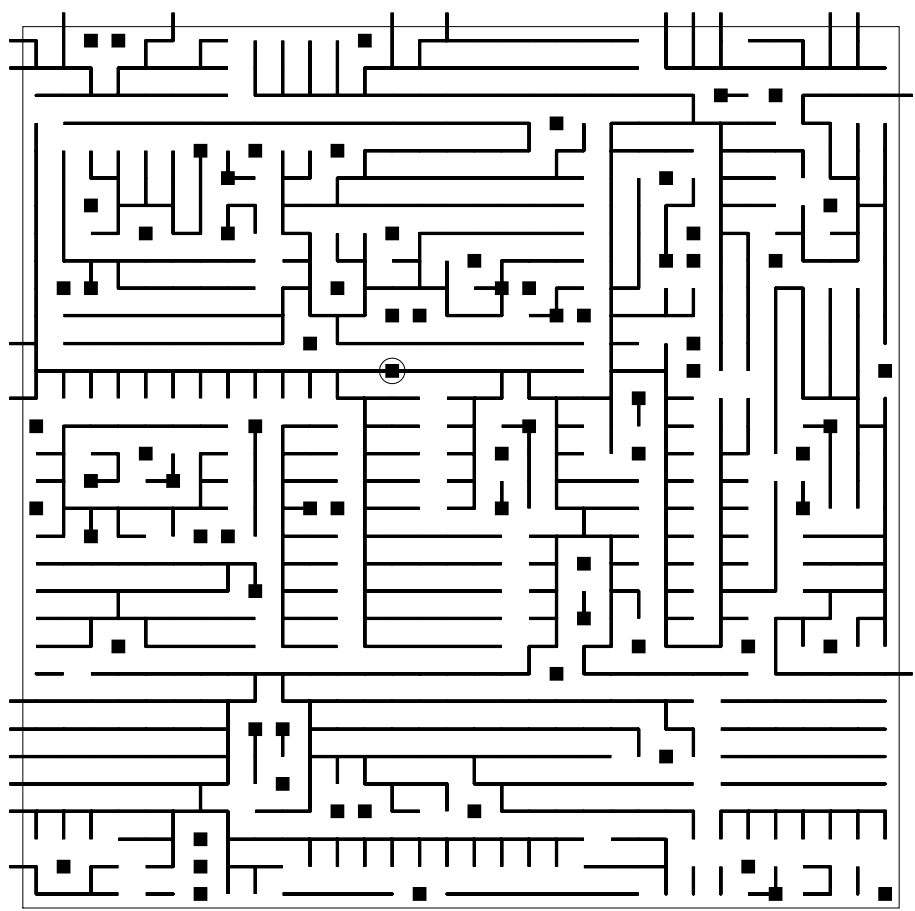
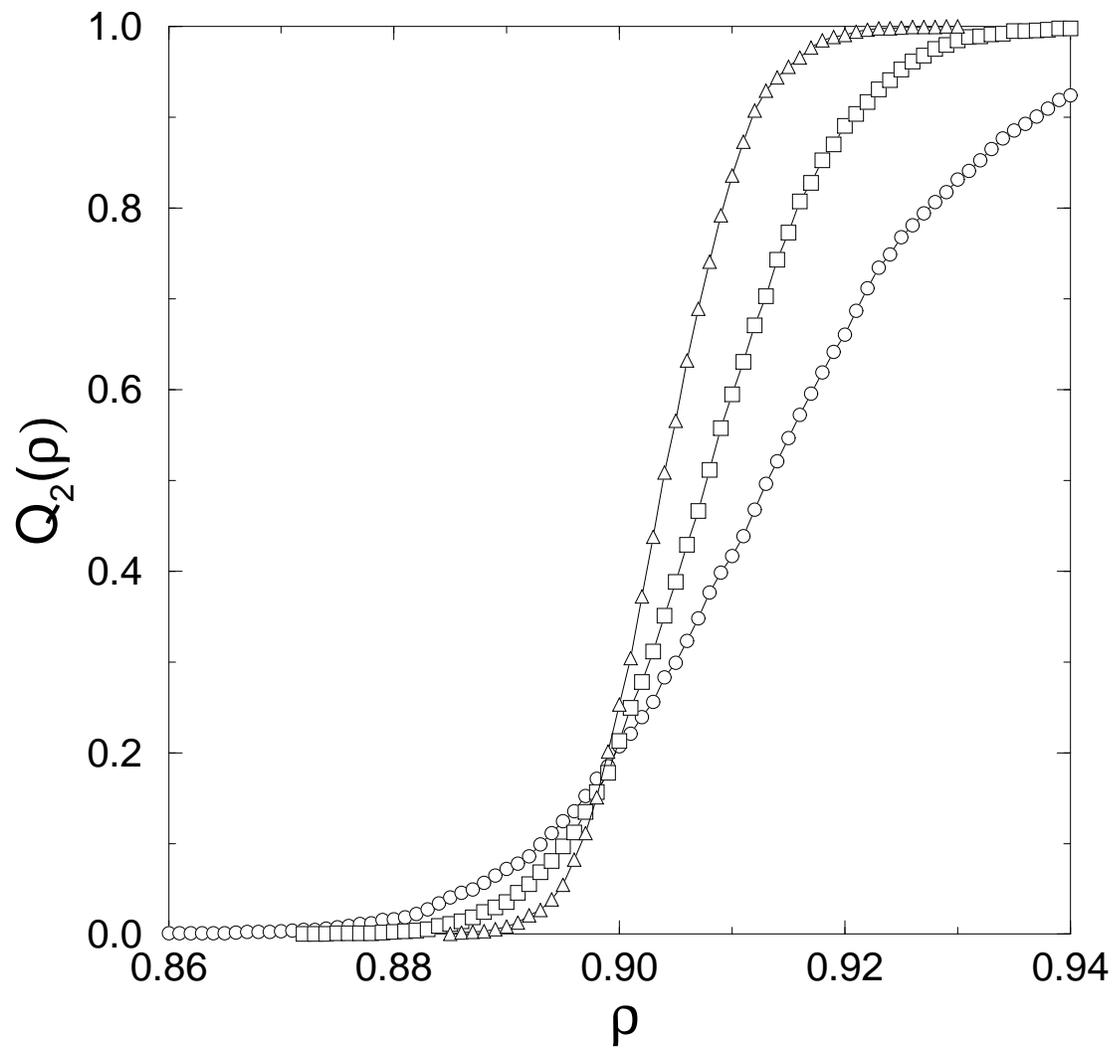
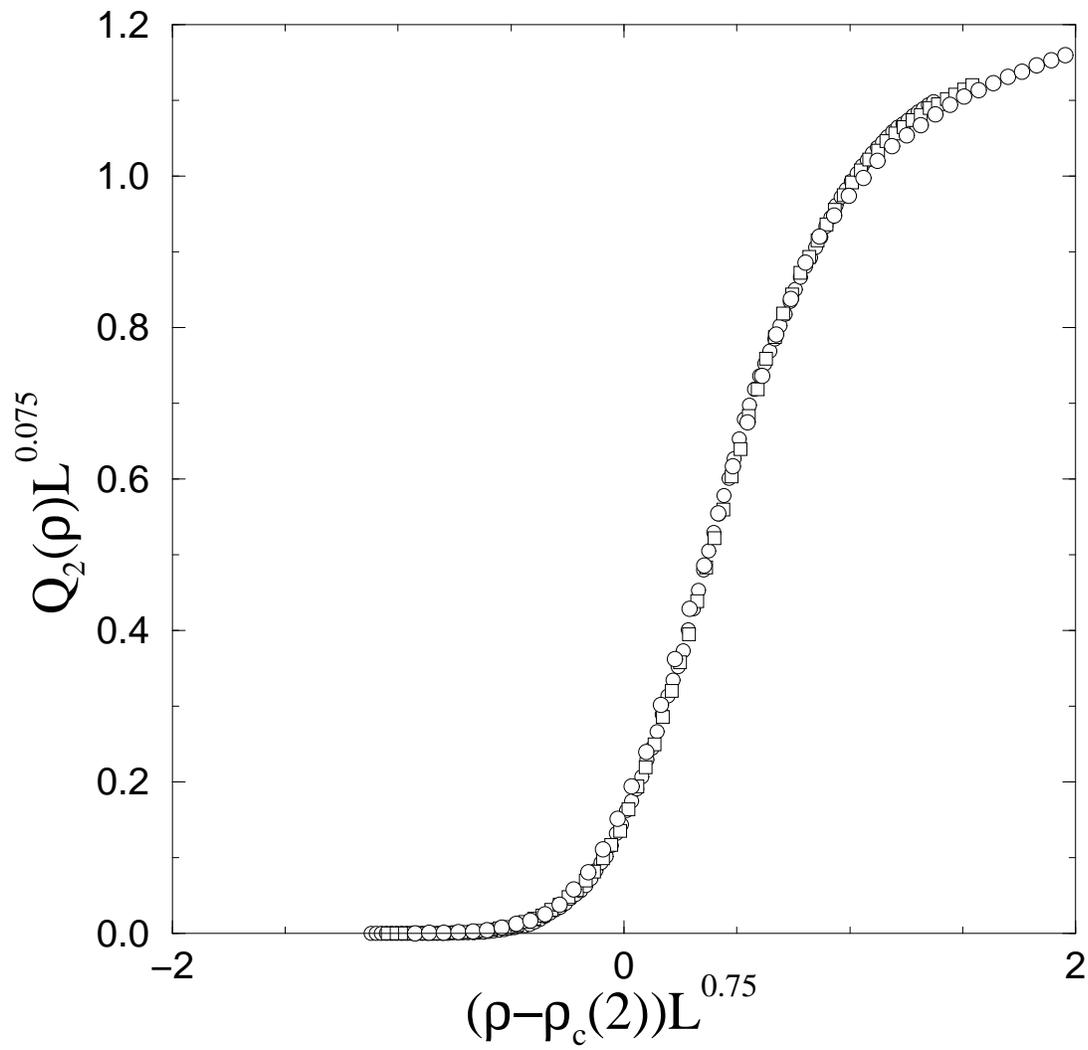
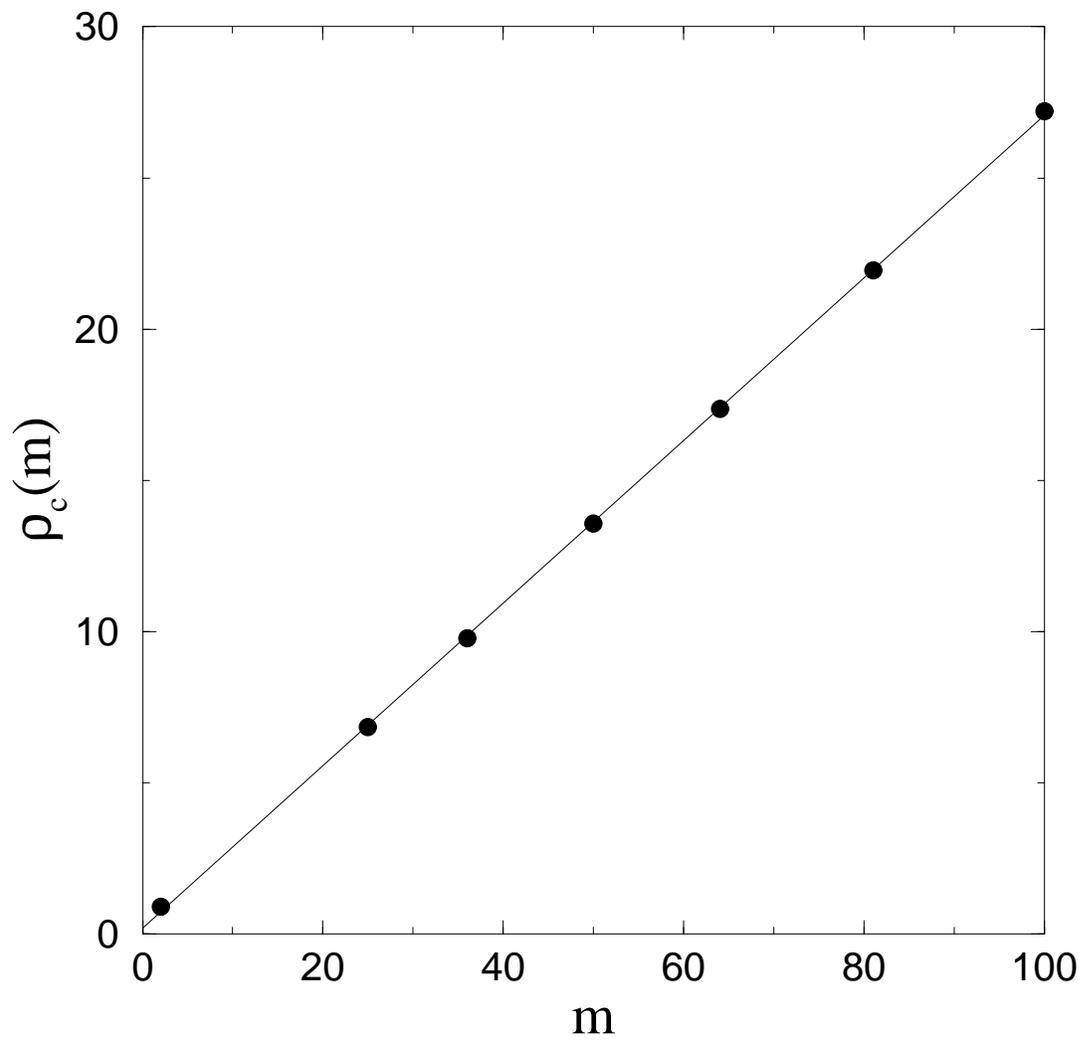
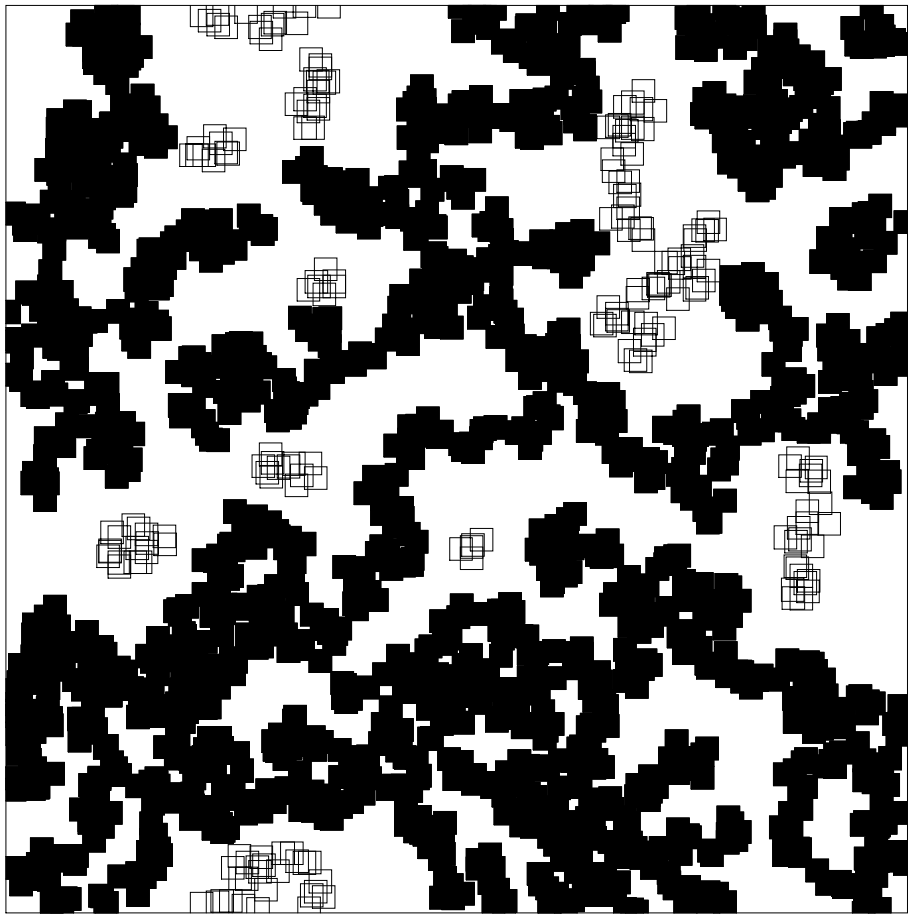
